\documentstyle[12pt]{article}
 
\input epsf 

   \def\unlock{\catcode`@=11}

   \unlock
 
   \def\gsim{\mathrel{\mathpalette\@versim>}}
 
   \def\@versim#1#2{\vcenter{\offinterlineskip
        \ialign{$\m@th#1\hfil##\hfil$\crcr#2\crcr\sim\crcr } }}

\begin{document}

\renewcommand{\thefootnote}{\alph{footnote}} 
\begin{titlepage}
 
\hspace*{\fill}\parbox[t]{2.8cm}{DESY 94-128 \\ SCIPP 94/20 \\ August 1994}
 
\vspace*{1cm}
 
\begin{center}
\large\bf
Azimuthal Angle Decorrelation in Large-rapidity Dijet Production
at the Tevatron\footnote{Invited talk presented by V.D.D. at the ``QCD 94'' 
workshop, Montpellier, France, July 7-13, 1994}
\end{center}
 
\vspace*{0.5cm}
 
\begin{center}
Vittorio Del Duca \\
Deutsches Elektronen-Synchrotron \\
DESY, D-22607 Hamburg , GERMANY\\
\vspace*{0.5cm}
and\\
\vspace*{0.5cm}
Carl R. Schmidt \footnote{Supported in part by the U.S.
Department of Energy.} \\
Santa Cruz Institute for Particle Physics\\
University of California, Santa Cruz, CA 95064, USA
\end{center}
 
\vspace*{0.5cm}
 
\begin{center}
\bf Abstract
\end{center}
 
\noindent
In this talk we examine the azimuthal angle decorrelation in dijet 
production at large rapidity intervals at the Tevatron as a possible
signature of the BFKL evolution.
\end{titlepage}
 
\baselineskip=0.8cm
 
\section{Introduction}

The state-of-the-art in jet physics at hadron colliders is described by
next-to-leading-order (NLO) QCD parton-level calculations\cite{EKS}, 
\cite{EKS2}, \cite{GGK}. They appear to be in very good
agreement with the one- and two-jet inclusive distributions obtained
from the data of the CDF experiment at the Fermilab Tevatron
Collider\cite{CDF}, \cite{Abe}.
However, since data are being collected at the CDF and D0 detectors at
larger and larger rapidities, it may be possible to probe kinematic
configurations where this fixed-order analysis is inadequate.
This could occur when the cross section contains logarithms
of large ratios of kinematic invariants.  Typical invariants are
the hadron-hadron center-of-mass energy $\sqrt{s}$, the parton-parton
center-of-mass energy $\sqrt{\hat s}=\sqrt{x_Ax_Bs}$, where $x_A$ and $x_B$
are the momentum fractions of the partons originating the hard scattering,
and the momentum transfer $Q$, which is of the order of the transverse 
momentum of the jets produced in the hard scattering. If $x_A$ and $x_B$
are large, then the large logarithms, $\ln(\hat s/Q^2)$, factorize
into the partonic subprocess cross section. These
logarithms, which are of the size of the rapidity interval in the 
scattering process, can be resummed by using the techniques of Balitsky,
Fadin, Kuraev, and Lipatov (BFKL)\cite{BFKL}. 
 
In analyzing dijet production
experimentally so that it most closely resembles the configuration
assumed in the BFKL theory, the jets are ordered first by their
rapidity rather than by their energy\cite{MN}.
Thus, we look at all the jets in the event
that are above a transverse momentum cutoff $p_{\perp min}$,
using some jet-definition algorithm, and rank them by their rapidity.
We then tag the two jets with the largest and smallest rapidity and
observe the distributions as a function of these two {\it tagging jets}.
The cross section is inclusive so that the distributions are affected by
the hadronic activity in the rapidity interval $y$ between the tagging
jets, whether or not these hadrons pass the jet-selection criteria.
We will refer to these hadrons in the rapidity interval as
{\it minijets}. For dijet events at large rapidity intervals the BFKL theory
systematically resums the leading powers in the
rapidity interval $y$, including both real and virtual gluon corrections.
 
In ref.~\cite{DDS} we showed that the exponential enhancement with
$y$ in dijet production at fixed $x_A$ and $x_B$ ,
originally suggested as a signature of the BFKL minijets by
Mueller and Navelet\cite{MN}, is highly suppressed by the parton distribution
functions at Tevatron energies.
However, we found that other observables such as the transverse momentum 
$p_\perp$ distribution and the jet-jet correlations in $p_\perp$ and
azimuthal angle $\phi$ are significantly affected by the minijet resummation.
For example, we saw that these correlations are not a leading
feature of the expansion in the rapidity interval. Accordingly, they fade away
as the rapidity interval increases.

However, in comparing the BFKL resummation with the exact 
${\cal O}(\alpha_s^3)$ calculation\cite{DDS2}, we noticed that the 
large-rapidity approximation to the kinematics
causes a serious error in the BFKL predictions
when the tagging jets are not back-to-back in $p_{\perp}$ and $\phi$.  In
order to account for this error we introduced an effective rapidity $\hat y$
which restricts the phase space of the minijets in such a way that the 
truncation of the BFKL resummation to ${\cal O}(\alpha_s^3)$ agrees with the 
exact $2\rightarrow3$ ${\cal O}(\alpha_s^3)$ calculation. For large $y$,
the difference $y\!-\!\hat y$ is nonleading. Since the rapidity
variable which is resummed by BFKL is only defined up to transformations 
$y\rightarrow y+X$ where $X$ is subleading at large rapidities, we used 
$\hat y$ instead of $y$ in the BFKL resummation in order to obtain 
quantitatively more reliable predictions of the transverse momentum 
distributions. We use here the effective rapidity $\hat y$ in the BFKL
resummation to analyze the moments\cite{stir} of the
decorrelation in azimuthal angle. 

\section{The minijet resummation and the effective rapidity interval}

We are interested in the semi-inclusive production of two jets in 
hadron-hadron collisions $p_A\bar p_B \rightarrow j_1 j_2 + X$.
We describe the two partonic tagging jets by their transverse momenta 
and rapidities $(\vec{p}_{1\perp}, y_1)$ and $(\vec{p}_{2\perp},y_2)$, where
we always take $y_1>y_2$. We reexpress the jet rapidities in terms of the
rapidity interval $y=y_1-y_2$ and the rapidity boost ${\bar y}=(y_1+y_2)/2$.
This is convenient since we are mainly interested in the behavior of the
parton subprocess, which does not depend on $\bar y$.
For large values of $y$ the cross section for this process can be written
\begin{equation}
{d\sigma_0\over dy\,d{\bar y}\,dp_{1\perp}^2 dp_{2\perp}^2 d\phi}\,=\,
x^0_Ax^0_B\,f_{\rm eff}(x^0_A,\mu^2)f_{\rm eff}(x^0_B,\mu^2)\,
{d\hat\sigma_{gg}\over dp_{1\perp}^2 dp_{2\perp}^2 d\phi}\ ,
\label{general}
\end{equation}
where the parton momentum fractions are dominated by the contribution
from the two tagging jets
\begin{eqnarray}
x^0_A &=& {p_{1\perp} e^{y_1} \over\sqrt{s}}\nonumber\\
x^0_B &=& {p_{2\perp} e^{-y_2} \over\sqrt{s}},\label{pmf}
\end{eqnarray}
and $\mu$ is the factorization/renormalization scale.
In this limit the amplitude is dominated by $gg$, $qg$, and $qq$
scattering diagrams with gluon-exchange in the $t$-channel.  The
relative magnitude of the different subprocesses is fixed by the color
strength of the respective jet-production vertices, so it
suffices to consider only $gg$ scattering and to include the other
subprocesses by means of the effective parton distribution 
function $f_{\rm eff}(x,\mu^2)$ \cite{CM},\cite{MN}.
 
The higher-order corrections to the $gg$ subprocess cross section in
(\ref{general}) can be expressed via the solution of the BFKL
equation\cite{BFKL}, which is an all-order resummation in
$\alpha_s$ of the leading powers of the rapidity interval
\begin{equation}
{d\hat\sigma_{gg}\over dp_{1\perp}^2 dp_{2\perp}^2 d\phi}\,=\, 
{C_A^2\alpha_s^2 \over 4\pi p_{1\perp}^3 \, p_{2\perp}^3} \,
\sum_n e^{in(\phi-\pi)} M_n(y, p_{1\perp}, p_{2\perp}) \, 
\label{mini}
\end{equation}
with
\begin{equation}
M_n(y, p_{1\perp}, p_{2\perp})\,=\, 
\int_0^{\infty} d\nu e^{\omega(n,\nu)\, y}
\cos\left(\nu \, \ln{p_{1\perp}^2 \over p_{2\perp}^2} \right) 
\label{coff}
\end{equation}
and
\begin{equation}
\omega(n,\nu) = {2 C_A \alpha_s \over\pi} \bigl[ \psi(1) -
 {\rm Re}\,\psi ({|n|+1\over 2} +i\nu) \bigr],
\label{eigen}
\end{equation}
and $\psi$ the logarithmic derivative of the Gamma function.
Eq.~(\ref{mini}) can be expanded order by order in $\alpha_s$. 
At ${\cal O}(\alpha_s)$ and for
$p_{1\perp} \ne p_{2\perp}$ we obtain
\begin{equation}
{d\hat\sigma_{gg}^{(1)}\over dp_{1\perp}^2 dp_{2\perp}^2 d\phi}\, =
{C_A^2\alpha_s^2
\over 4\pi p_{1\perp}^2 \, p_{2\perp}^2}\ {C_A\alpha_s y \over
p_{1\perp}^2 + p_{2\perp}^2 +2p_{1\perp}p_{2\perp}\cos{\phi}}. 
\label{explic}
\end{equation}
This can be compared with a
fixed-order calculation of dijet production at the same order of $\alpha_s$,
computed through the 2$\rightarrow$3 parton amplitudes,
\begin{equation}
{d\sigma\over dy\,d{\bar y}\,dp_{1\perp}^2\,dp_{2\perp}^2\,d\phi}\, =\,
\int_{y_2}^{y_1}dy_3\int \sum_{ij} x_A x_B 
f_{i/A}(x_A,\mu^2) f_{j/B}(x_B,\mu^2)\,
{d\hat\sigma_{ij}\over dp_{1\perp}^2 dp_{2\perp}^2 dy_3 d\phi}\ , 
\label{excross}
\end{equation}
where $y_3$ is the rapidity of the third final-state parton, integrated over
the interval spanned by the tagging jets. 
$f_{i(j)} = Q,\bar Q, G$ labels 
the distribution function of the parton species and flavor 
$i$($j$) = $q,\bar q, g$ inside hadron $A$($B$).
We include all parton subprocesses\cite{exact},
and use the exact values of the parton momentum fractions
\begin{eqnarray}
x_A &=& {p_{1\perp} e^{y_1} + p_{2\perp} e^{y_2} + p_{3\perp} e^{y_3}\over
\sqrt{s}}\nonumber\\
x_B &=& {p_{1\perp} e^{-y_1} + p_{2\perp} e^{-y_2} + p_{3\perp} e^{-y_3}
\over\sqrt{s}}.\label{expmf}
\end{eqnarray}
In order to avoid the collinear singularity between final-state partons,
which is however subleading at large rapidities\cite{DDS2}, configurations
in (\ref{excross}) where the distance $R$ between two of the partons on the
Lego plot in azimuthal angle and rapidity is smaller than the jet cone 
size $R_{cut}$ are discarded. In the large-$y$ limit (\ref{excross})
reduces to (\ref{general}), with the parton cross section given by 
(\ref{explic}). 

However, we have seen in ref.\cite{DDS2} that the large-$y$ 
approximation seriously overestimates the cross section when the two tagging
jets are not back-to-back in $p_{\perp}$ and $\phi$, even for rapidity 
intervals as large as $y=6$.
This occurs because the large-$y$ cross section (\ref{explic}) assumes that
the third (minijet) parton can be produced anywhere within the rapidity
interval $[y_2,y_1]$ with equal probability, whereas in the full
$2\rightarrow3$ cross section the probability is highly suppressed by the
structure functions when the third jet strays too far from the center of this
interval. In ref.\cite{DDS2} we introduced an ``effective'' rapidity interval 
$\hat y$ to take into account the fact
that the range in rapidity spanned by the minijets is
typically less than the kinematic rapidity interval $y$. We define 
$\hat y(n,p_{1\perp},p_{2\perp},\bar y, y)$ by
\begin{equation}
\hat y\ \equiv\ y\ {
\displaystyle \int d\phi\, \cos(n\phi)\ \big(d\sigma/dyd\bar y
dp_{1\perp}dp_{2\perp}d\phi\big) 
\over
\displaystyle\int d\phi\, \cos(n\phi)\ \big(
d\sigma_0/dyd\bar ydp_{1\perp}dp_{2\perp}d\phi\big)},\label{effecty}
\end{equation}
where $n$ is the Fourier series index of eq.~(\ref{mini}).
The cross section in the numerator is that of eq.~(\ref{excross}) and is 
computed 
using the exact kinematics (\ref{expmf}), while the cross section in the 
denominator is that of eq.~(\ref{general}) and is computed using the large-$y$ 
kinematics (\ref{pmf}).  The denominator can easily
be computed analytically using the large-$y$ solution
(\ref{explic}).  Note that $\hat
y$ is defined so that if we replace $y\rightarrow\hat y$ in the BFKL solution
(\ref{mini}) and truncate to ${\cal O}(\alpha_s^3)$ we recover the exact
$2\rightarrow3$ cross section. 

\section{The azimuthal angle decorrelation}

Our aim is now to reconsider the azimuthal angle 
decorrelation\cite{DDS},\cite{stir} by using the effective rapidity 
$\hat y$ instead of
the kinematic one $y$ in the BFKL resummation. Using $\hat y$, we expect that 
the distribution will be more populated around the peak $\phi = \pi$ where the
tagging jets are back-to-back, and less populated on its tails where the
tagging jets are decorrelated, as compared
to the $\phi$ decorrelation plot we considered in ref.\cite{DDS}. This is 
because the larger the decorrelation, the larger the overestimating error 
in approximating the parton momentum fractions with eq.~(\ref{pmf}).
However, in practice the computation of the Fourier coefficients (\ref{coff}) 
using eq.~(\ref{effecty}) for $\hat y$ becomes quickly very time-consuming 
as $n$ grows. 
To test the BFKL picture, it is then more convenient to
compute the moments of the azimuthal angle\cite{stir}, defined by
\begin{equation}
<\cos n(\phi\!-\!\pi)>\, = 
{\int_0^{2\pi} d\phi \cos n(\phi\!-\!\pi)\, 
\big(d\sigma_0/dy d\phi\big)
\over \int_0^{2\pi} d\phi \big(d\sigma_0/dy d\phi\big)}. 
\end{equation}
For a $\delta$-function distribution at $\phi=\pi$, as occurs at the Born 
level, all of the moments will equal one, while for a flat distribution all
of the moments will equal zero for $n\ge1$.  Thus, the decay of the moments 
from unity is a good measure of the decorrelation in $\phi$.

In Fig.~1 we plot in the solid curves
the first two moments of the azimuthal angle as a function 
of $y$, calculated using the effective rapidity $\hat y$ in the BFKL solution.
We use the LO
CTEQ2 parton distribution functions\cite{cteq} with the ren./fact.
scales set to $\mu^2=p_{1\perp}p_{2\perp}$. 
To facilitate the comparison with the experimental data we choose the
kinematic parameters of the data analysis of the 1993 run at the D0 detector.
Namely, we require that one of the tagging jets must have 
$p_{\perp}>50$~GeV, while the other must have $p_{\perp}>20$~GeV.
The rapidity boost $\bar y$ is integrated over, subject to the constraint
$|y_1|_{max}=|y_2|_{max}=3.2$.  The rapidity interval is integrated in unit
bins centered around the variable $y$ of the plot.  Finally, to compare with 
our previous analysis, we also show in the dashed curves the moments 
calculated using the kinematic rapidity.

\begin{figure}[htb]
\vspace{12pt}
\vskip-4cm
\epsfysize=16cm
\centerline{\epsffile{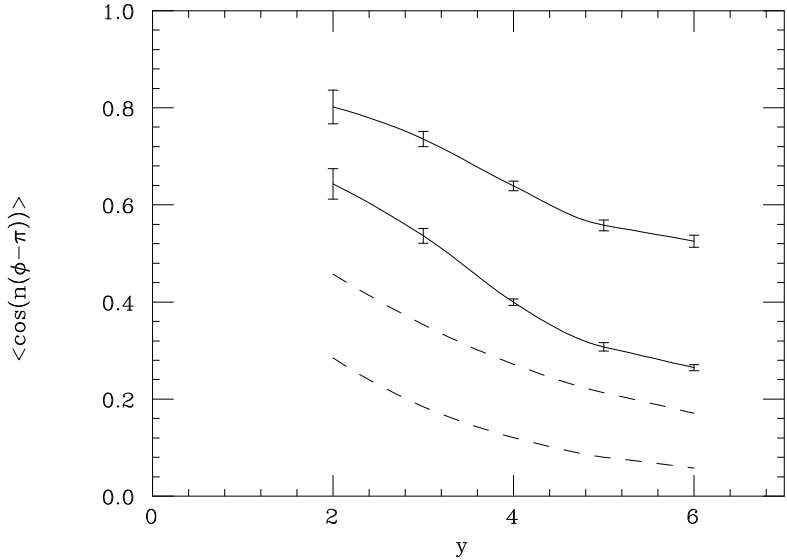}}
\vspace{18pt}
\vskip-5cm
\caption{The moments of the azimuthal angle distribution for $n=1$ and 2,
computed with the BFKL resummation using the effective rapidity $\hat y$
(solid) and using the kinematic rapidity $y$ (dashed).  
The error bars are  
the estimated statistical error from the monte carlo integration.}

\label{fig:momentos}
\vspace{12pt}
\end{figure}

As expected, the naive BFKL analysis (dashed) tends to overestimate the
decorrelation in azimuthal angle.  The first two moments are typically
two to three times smaller than those calculated using $\hat y$ in the
BFKL solution (solid).  However, there is still a 
substantial decorrelation exhibited in the solid curves which should be
experimentally measurable.
For small rapidities one would expect these results to be
invalidated by nonleading effects, but for larger rapidities 
(optimistically, perhaps $y\gsim4$) the minijet effects should begin to 
dominate.  This is a particularly nice process for probing the BFKL structure,
from both an experimental and a theoretical point of view.  Experimentally,
the $\phi$ measurements are much less affected by detector resolution than
the $p_{\perp}$ measurements \cite{terry}.  Theoretically, the 
azimuthal angle distribution is less sensitive to the 
factorization/renormalization scale used.  Thus, a measurement of the 
azimuthal angle decorrelation should be an excellent probe for the 
validity range of the BFKL dynamics.

\end{document}